\documentclass[%
 preprint,
superscriptaddress,
 amsmath,amssymb,
 aps,
]{revtex4-1}

\usepackage{graphicx}
\usepackage{dcolumn}
\usepackage{bm}
\usepackage{amsfonts}
\usepackage{hyperref}
\usepackage[mathlines]{lineno}

\begin{document}


\title{Investigating nonflow contribution subtraction in d-Au collisions with AMPT model}

\author{Zuman Zhang}\thanks{Email: zuman.zhang@hue.edu.cn}
\affiliation{School of Physics and Mechanical Electrical \& Engineering, Hubei University of Education, Wuhan 430205, China}
 \affiliation{Institute of Theoretical Physics, Hubei University of Education, Wuhan 430205, China}
 \affiliation{Key Laboratory of Quark and Lepton Physics (MOE), Central China Normal University, Wuhan 430079, China}
 \author{Sha Li}\thanks{Email: lisha@hue.edu.cn}
 \affiliation{School of Physics and Mechanical Electrical \& Engineering, Hubei University of Education, Wuhan 430205, China}
\author{Ning Yu}\thanks{Email: ning.yuchina@gmail.com}
\affiliation{School of Physics and Mechanical Electrical \& Engineering, Hubei University of Education, Wuhan 430205, China}
 \affiliation{Institute of Theoretical Physics, Hubei University of Education, Wuhan 430205, China}
 \affiliation{Key Laboratory of Quark and Lepton Physics (MOE), Central China Normal University, Wuhan 430079, China}
 \author{Qiao Wu}
 \affiliation{School of Physics and Mechanical Electrical \& Engineering, Hubei University of Education, Wuhan 430205, China}

\date{\today}

\begin{abstract}
This paper presents research that focuses on nonflow contribution subtraction in heavy-ion collisions, using a multiphase transport model (AMPT). Specifically, the study aims to investigate the behavior of charged particle elliptic flow ($v_{\rm 2}$) in d-Au collisions at a collision energy of $\sqrt{s_{\rm NN}} = 200$~GeV and to determine the impact of nonflow sources, such as jet correlations and resonance decays, in small collision systems. To reduce nonflow effects, the per-trigger yield distribution in peripheral d-Au collisions or pp collisions with the same collision energy is subtracted. Our results show that the nonflow effects in central and mid-central collisions are not strongly dependent on subtracting the per-trigger yield distribution in peripheral d-Au collisions or pp collisions. Furthermore, the elliptic flow of charged particles, after removing nonflow effects through two subtracting methods from this work, exhibits consistency in various collision centrality classes. We also discuss comparisons with measurements from d-Au collisions at $\sqrt{s_{\rm NN}} = 200$~GeV. Overall, this work provides valuable insights and serves as a reference for researchers studying nonflow contribution subtraction in experiments with small collision systems.

\begin{description}\item[PACS numbers]
\verb+ 25.75.Ld Collective flow, 24.10.Lx Monte Carlo+ simulations (including hadron and parton cascades and string breaking models)
\end{description}
\end{abstract}

\maketitle

\section{Introduction}
\label{sec:intro}
  In high-energy heavy-ion collision research, the goal is to understand the properties of the quark-gluon plasma (QGP), a state of matter characterized by high energy density and temperature~\cite{Muller:2006ee,Bazavov:2011nk}. One of the key observables in this field is the azimuthal anisotropy of final state particles, which provides information about the transport features of the QGP~\cite{Ollitrault:1992bk}.
  The elliptic flow ($v_{\rm 2}$) is a crucial observable in studying the collective motion of the QGP and is obtained through the Fourier expansion of the azimuthal distribution of emitted particles in transverse momentum space. The second-order coefficient in this expansion provides valuable insights into the behavior of the QGP and its properties.
  The study of elliptic flow is an important aspect of high-energy heavy-ion collision research as it helps to deepen our understanding of the quark-gluon plasma and its characteristics.

  In order to better understand the effects of cold nuclear matter on the interpretation of measurements in heavy-ion collisions, we conducted a study of small collision systems such as p(d)+A collisions. Our research focused on exploring key aspects of cold nuclear matter effects, such as the modification of parton distribution functions~\cite{Eskola:2009uj}, the broadening of  $k_{\rm T}$~\cite{Kopeliovich:2002yh}, and energy loss in cold nuclear matter~\cite{Kang:2014hha}.
  To our surprise, in small collision systems, using the two-particle azimuthal correlation method, high-multiplicity p+A collisions at $\sqrt{s_{\rm NN}} = 5.02$~TeV in the midrapidity region by the ALICE, ATLAS, and CMS collaborations ~\cite{CMS:2012qk,Abelev:2012ola,Aad:2012gla,ABELEV:2013wsa,Aaboud:2016yar} and at forward rapidity by the LHCb collaboration~\cite{Aaij:2015qcq}, we discovered long-range structures in two-particle correlations that were associated with a positive $v_{\rm 2}$ for hadrons.
  In addition, at lower beam energies, long-range correlations were also observed in d-Au~\cite{Adare:2013piz,Adare:2014keg,Adamczyk:2015xjc} and $^3$He-Au collisions~\cite{Adare:2015ctn} by the PHENIX and STAR collaborations at RHIC.

  To address nonflow effects, various strategies have been developed to eliminate correlations that are not associated with collectivity and arise from sources such as jet interactions, resonance decays, and so on. Typically, in small collision
  system experiments these nonflow contributions were suppressed by requiring a separation in pseudorapidity between paired particles or by subtracting correlations measured in low-multiplicity~\cite{Abelev:2012ola,Aad:2014lta} or pp collisions~\cite{Aad:2019lta}. The long-range correlations were then isolated using a standard template fit procedure~\cite{Aad:2019ajj}. In the experimental analysis, the contribution of nonflow has not been calculated by both methods for same collision system. So, we analyze the $p_{\rm T}$  distribution of anisotropic flow ($v_{\rm 2}$) in d-Au collisions at $\sqrt{s_{\rm NN}} = 200$~GeV, using the multiphase transport (AMPT) model~\cite{Lin:2004en} to investigate the nonflow contribution subtraction from both peripheral and pp collisions. The paper starts with a brief introduction to the AMPT model. Then, the method of nonflow contribution subtraction is detailed. Finally, the results are presented, along with a discussion and conclusion.

\section{Event Generation and Definition of anisotropic flow}
\label{sec:Generation_Methodology}
\subsection{A Multi-Phase Transport (AMPT) model}
\label{subsec:APMT}

  The AMPT model~\cite{Lin:2004en} is a hybrid transport model used to study collective behavior in heavy ion collisions. It consists of four components: initial conditions, partonic interactions, conversion from partonic to hadronic matter, and hadronic interactions.

  The Lund string fragmentation function is determined by the parameters $a$ and $b$  in HIJING~\cite{Wang:1991hta} as $f(z)\propto
  z^{-1}(1-z)^{a}\exp{(-bm^{2}_{\perp}/z)}$, where $z$ is the light-cone momentum fraction of the produced hadron of transverse mass $m_{\perp}$ with respect to the fragmenting string.
  The Zhang's Parton Cascade (ZPC) model~\cite{Zhang:1997ej}, which calculates parton-parton scattering using cross sections $\sigma\approx\frac{9\pi\alpha_{s}^{2}}{2\mu^{2}}$  based on a Debye screening mass $\mu$, is used to simulate the evolution of the partonic phase.
  After the partons stop interacting in ZPC, the quarks are subjected to a hadronization process based on the quark coalescence model, which combines the nearest quarks in coordinate space into hadrons.
  The hadrons formed during quark coalescence are then subjected to hadronic stage evolution, which is handled by a relativistic transport (ART) model~\cite{Li:1995pra} with the input cross section for various hadron-hadron scattering channels.

  In this work we use the program version with string melting AMPT model. We have conducted approximately 10 million  AMPT events for d-Au collisions at $\sqrt{s_{\rm NN}} = 200$~GeV and pp collisions at $\sqrt{s} = 200$~GeV.
  The event centrality in $d$$+$Au is determined by impact parameter $b$ from AMPT events.
  We use central, mid-central and peripheral event samples comprising the top 5\%, 10--20\%,  20--30\%, 30--40\%, 40--50\%
   and 50--88\% collision centrality intervals of the total charge particle elliptic flow ($v_{\rm 2}$) distributions, respectively.

\subsection{Definition of anisotropic flow}
\label{subsec:epmethod}

  Typically, the magnitude of azimuthal anisotropies is quantified using a Fourier decomposition of the particle azimuthal distribution given by
 \begin{equation}
 {{\rm d}^2N \over {{\rm d}p_{\rm T}{\rm d}\varphi }} = {1\over 2\pi}
 {{\rm d}N \over {\rm d} p_{\rm T}}
 \bigg ( 1 + 2 \sum_{{n} = 1}^{\infty} v_{n} (p_{\rm T}) {\rm cos}
 \lbrack {n} ( \varphi - \Psi_{n} ) \rbrack \bigg),
 \label{eq:azimuthal_distribution}
 \end{equation}
  where $\varphi$ and $p_{\rm T}$ are the particle azimuthal angle and transverse momentum, respectively.
  The anisotropy of produced particles is defined by the Fourier coefficients $v_{n}$~\cite{Voloshin:1994mz}
  and the azimuthal angle of the symmetry plane for the $n^{\rm th}$ harmonic is denoted by $\Psi_{n}$.
  The largest contribution to the asymmetry of collisions is provided by the second Fourier coefficient $v_2$ referred to as elliptic flow~\cite{Ollitrault:1992bk,Voloshin:1994mz}.

\section{Nonflow contribution subtraction method}
\label{sec:2correl}

 The method using two-particle correlations to extract the azimuthal anisotropy is extensively discussed in Refs.~\cite{CMS:2012qk,Abelev:2012ola,Aad:2012gla,ABELEV:2013wsa,Aaboud:2016yar,Aaij:2015qcq,Adam:2015bka,Acharya:2017tfn,Acharya:2018dxy}.
 The correlation between two particles (denoted trigger and associated particle) is measured
 as a function of the azimuthal angle difference $\Delta \phi$ (defined within $-\pi/2$ and $3\pi/2$) and pseudorapidity
 difference $\Delta \eta$. While the trigger particles are charged particles, the analysis is done for charged associated particles (denoted $h-h$). In this work, we follow the analysis in experiments at the RHIC energy~\cite{Adare:2013piz}. In AMPT events, charged hadrons with $0.5<p_T<3.5$~GeV/$c$ are used, each pair includes at least one particle at low $p_T$
($0.5<p_{T}<0.75$~GeV/$c$).  To minimize the contribution from small-angle
 correlations pairs are restricted to pseudorapidity
 separations of $0.48<|\Delta\eta|<0.7$.
 The correlation is expressed in terms of $Y$, the associated yield per trigger particle defined as:
 \begin{equation}
 Y = \frac{1}{N_{\rm trig}} \frac{{\rm d}N_{\rm assoc}}{{\rm d}\Delta \phi},
 \label{eq:Y}
 \end{equation}
 where $N_{\rm trig}$ is the total number of trigger particles in the event class and $p_{\rm T}$ interval, $N_{\rm assoc}$ is the total number of associatied particles in the event class and $p_{\rm T}$ interval.

 We use the zero-yield-at-minimum (ZYAM)
 method~\cite{Ajitanand:2005jj}, where one assumes that the number of correlated pairs is zero at the correlation function minimum.
 This background contribution is
 obtained for the central, mid-central, peripheral and pp collisions samples by
 performing fits to the conditional yields using a functional form
 composed of a constant pedestal and two Gaussian peaks, centered at
 $\Delta \phi=0$ and $\pi$. The minimum of this function,
 $b_{\rm ZYAM}$, is subtracted from the conditional yields,
 and the result is:
 $Y(\Delta\phi) = \frac{1}{N_{\rm trig}}\frac{dN_{\rm assoc}}{d\Delta\phi} - b_{\rm ZYAM}$.
 The conditional yields
 $Y_{\rm c}(\Delta \phi)$, $Y_{\rm mc}(\Delta \phi)$, $Y_{\rm p}(\Delta \phi)$ and $Y_{\rm pp}(\Delta \phi)$ are related to central, mid-central, peripheral and pp collisions events,
 respectively. Their
 difference are $\Delta Y_{\rm cp} (\Delta \phi) = Y_{\rm c}(\Delta \phi) - Y_{\rm p}(\Delta \phi)$, $\Delta Y_{\rm mcp} (\Delta \phi) = Y_{\rm mc}(\Delta \phi) - Y_{\rm p}(\Delta \phi)$, $\Delta Y_{\rm cpp} (\Delta \phi) = Y_{\rm c}(\Delta \phi) - Y_{\rm pp}(\Delta \phi)$ and $\Delta Y_{\rm mcpp} (\Delta \phi) = Y_{\rm mc}(\Delta \phi) - Y_{\rm pp}(\Delta \phi)$ are associated with
 subtracting the per-trigger yield distribution in peripheral d-Au collisions or pp collisions, respectively. These subtraction removes any centrality independent correlations,
 such as effects from jet correlations, resonance decays, and so on.

 Fourier coefficients can be extracted from the $\Delta \phi$ projection of the per-trigger yield by
 a fit with:
 \begin{equation}
 \frac{1}{N_{\rm trig}}
 \frac{{\rm d}N_{\rm assoc}}{{\rm d} \Delta \phi} = a_0 + \sum_{ {\rm n } = 1}^{3} 2 a_{\rm n} {\rm  cos} ({\rm n} \Delta\phi)
 \label{eq:fourier}
 \end{equation}

  To quantify the relative amplitude of the azimuthal modulation,
  we define
 \begin{equation}
 c_{n} = a_{n} / \left( b^{c}_{\rm ZYAM} + a_{0} \right),
 \label{eq:cn}
 \end{equation}
 where $b^c_{\rm ZYAM}$ is $b_{\rm ZYAM}$ in central events~\cite{Ajitanand:2005jj}.

 The method using two-particle correlations to the $v_{n}^{h}\{{\rm 2PC}\}$ coefficient of order $n$ for a particle $h$ is defined as~\cite{ABELEV:2013wsa}:

 \begin{equation}
 v_{n}^{h}\{{\rm 2PC}\} = \sqrt{c_{n}^{h-h}}
 \label{eq:v2}
 \end{equation}

\section{Results and Discussions}
\label{sec:results}

\begin{figure}
\includegraphics[width=0.45\textwidth]{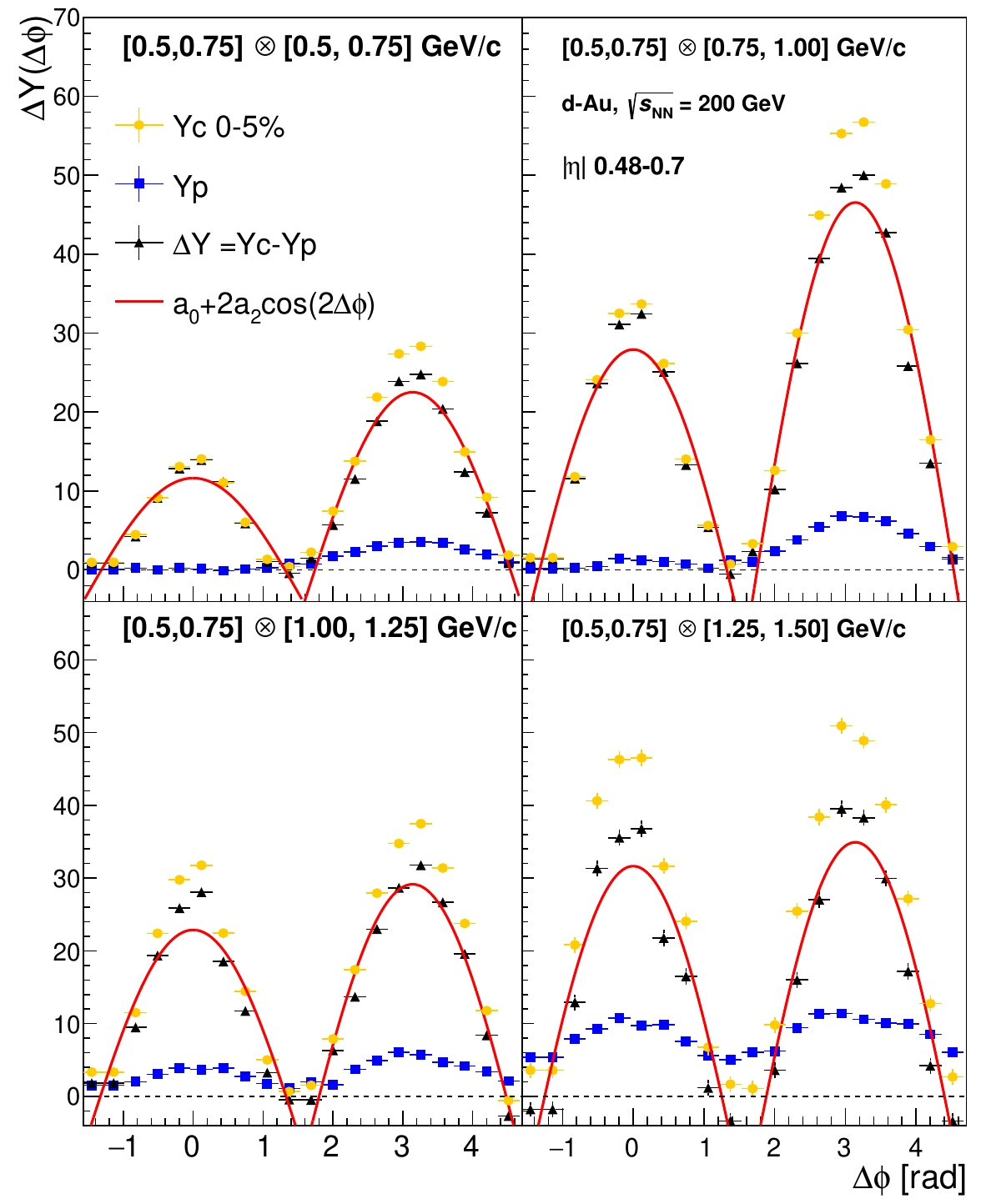}
\caption{Charged particles azimuthal conditional yields $Y\left(\Delta\phi\right)$
for 0\%--5\% most central ($Y_{\rm c}(\Delta \phi)$, orange circles) and
 50\%--85\% peripheral ($Y_{\rm p}(\Delta \phi)$, blue squares) collisions with $0.48<|\Delta\eta|<0.7$.
Difference $\Delta$$Y\left(\Delta\phi\right)$  between most central and peripheral d-Au collisions ($\Delta Y_{\rm cp} (\Delta \phi) = Y_{\rm c}(\Delta \phi) - Y_{\rm p}(\Delta \phi)$, black triangle-ups), which is
 fit to $a_0 + 2a_2 \cos(2\Delta\phi)$(red curve), where
$a_0$ and $a_2$ are computed from the AMPT events.
}
\label{fig:Azimuthal_conditional_yields_1}
\end{figure}

\begin{figure}
\includegraphics[width=0.45\textwidth]{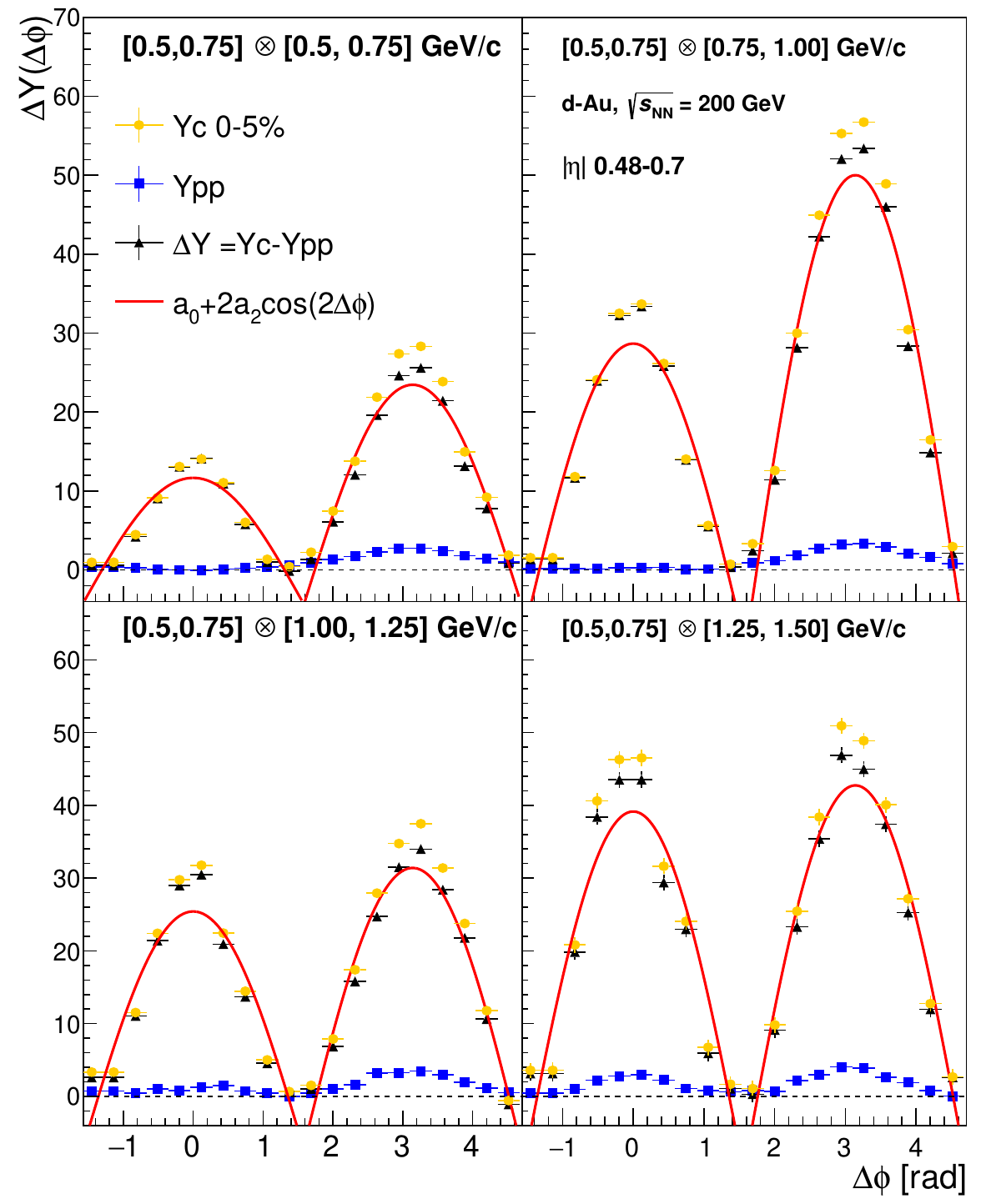}
\caption{Charged particles azimuthal conditional yields $Y\left(\Delta\phi\right)$
for 0\%--5\% most central ($Y_{\rm c}(\Delta \phi)$, orange circles) and
 pp ($Y_{\rm pp}(\Delta \phi)$, blue squares) collisions with $0.48<|\Delta\eta|<0.7$.
Difference $\Delta$$Y\left(\Delta\phi\right)$  between most central d-Au collisions and pp collisions ($\Delta Y_{\rm cpp} (\Delta \phi) = Y_{\rm c}(\Delta \phi) - Y_{\rm pp}(\Delta \phi)$, black triangle-ups), which is
 fit to $a_0 + 2a_2 \cos(2\Delta\phi)$(red curve), where
$a_0$ and $a_2$ are computed from the AMPT events.
}
\label{fig:Azimuthal_conditional_yields_2}
\end{figure}

  From Eq.~(\ref{eq:Y}), the charged particles conditional yields
  $Y_{\rm c}(\Delta \phi)$, $Y_{\rm p}(\Delta \phi)$ and $Y_{\rm pp}(\Delta \phi)$(0\%--5\% most central, peripheral and pp collisions events,
  respectively) are shown in Fig.~\ref{fig:Azimuthal_conditional_yields_1} and Fig.~\ref{fig:Azimuthal_conditional_yields_2} , along with their
  difference $\Delta Y_{\rm cp} (\Delta \phi) = Y_{\rm c}(\Delta \phi) - Y_{\rm p}(\Delta \phi)$ and $\Delta Y_{\rm cpp} (\Delta \phi) = Y_{\rm c}(\Delta \phi) - Y_{\rm pp}(\Delta \phi)$ are expressed as
  subtracting the per-trigger yield distribution in peripheral d-Au collisions at $\sqrt{s_{\rm NN}} = 200$~GeV or pp collisions at $\sqrt{s} = 200$~GeV, respectively.
  It is worth noting that any signal in the peripheral events and pp events is subtracted from the signal in the central events.
  For $\Delta\phi$ near 0 and $\pi$, $Y_c(\Delta\phi)$ is significantly larger than $Y_{\rm p}(\Delta\phi)$ and $Y_{\rm pp}(\Delta\phi)$.

  We discover that the distinction with 0--5\% most central, 50--85\% peripheral collisions and pp collisions are well described by the Eq.~(\ref{eq:fourier})
  as demonstrated in Fig.~\ref{fig:Azimuthal_conditional_yields_1} and Fig.~\ref{fig:Azimuthal_conditional_yields_2}. The charged particles coefficients $a_{n}$ are computed from the $\Delta
  Y(\Delta\phi)$ distributions as:
  $a_{n} = \langle \Delta Y(\Delta\phi) \cos(n\Delta\phi) \rangle$.
  The bracket $\langle...\rangle$ denotes an average over particles in the event sample with $\Delta\phi$.

\begin{figure}
\includegraphics[width=0.45\textwidth]{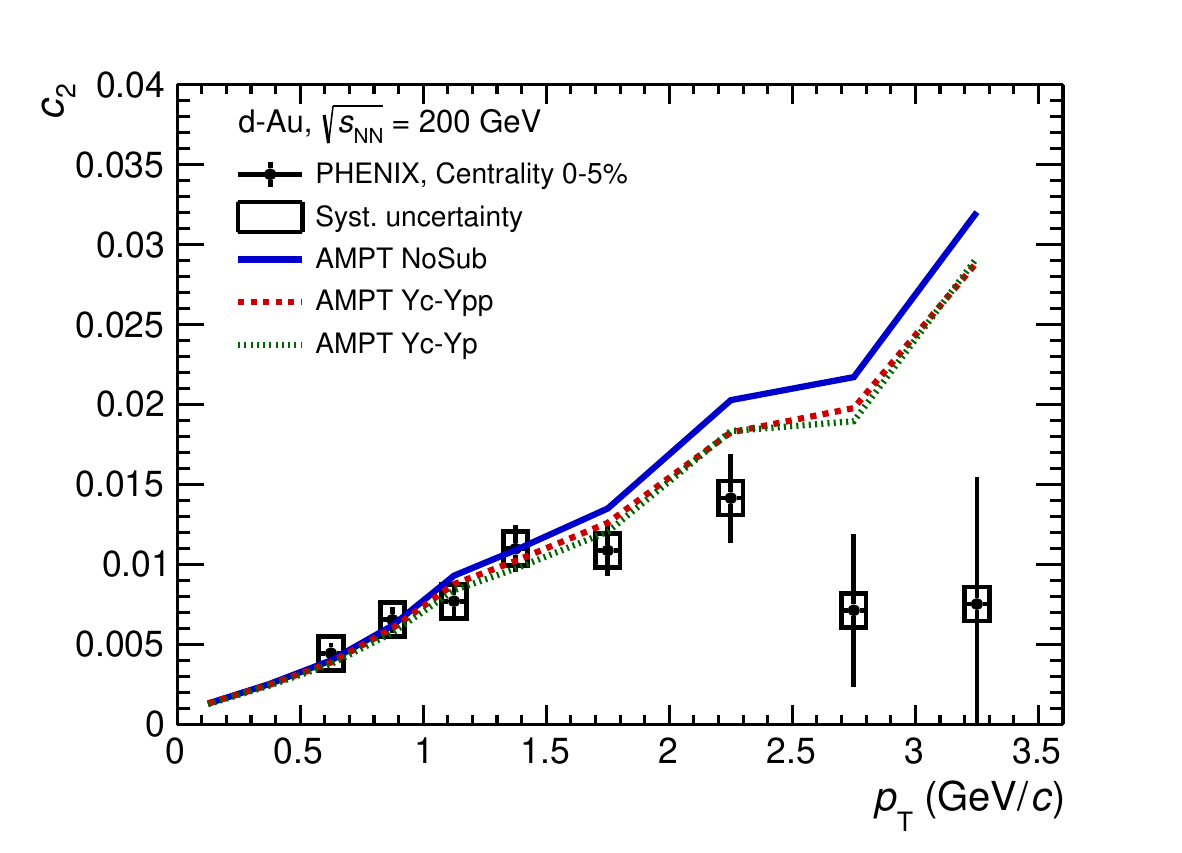}
\caption{With
$0.5<p^{t}_{T}<0.75$~GeV/$c$, $0.48<|\Delta\eta|<0.7$, the charged particles $2$th-order pair anisotropy $c_{2}$ of the 0--5\% most central collision excess as a function of associated particle $p_{\rm T}$.
The blue solid line is for $c_{2}$ of the 0--5\% most central d-Au collisions with nonflow effects.
The nonflow effects for 0--5\% most central d-Au collisions
reduced by subtracting the per-trigger yield distribution
in peripheral d-Au collisions (green dotted line) or pp collisions (red dashed line).
The $c_{2}$ of the central d-Au collision in the data is
also shown (black circles) from Ref.~\cite{Adare:2013piz}.
}
\label{fig:Azimuthal_conditional_yields_3}
\end{figure}

  From Eq.~(\ref{eq:cn}),charged particles $c_2$ is shown as a function of associated $p_{\rm T}$ in Fig.~\ref{fig:Azimuthal_conditional_yields_3} for central (0--5\%) d-Au collisions. With the nonflow effect, the blue solid line is higher than red dashed line or green dotted line (without the nonflow effect). The nonflow effects for 0--5\% most central d-Au collisions
  reduced by subtracting the per-trigger yield distribution
  in peripheral d-Au collisions (green dotted line) or pp collisions (red dashed line) are very similar. The charged particles $c_2$ from AMPT events and data are consistent from 0.5 to 2.0~GeV/$c$, then $c_2$ from AMPT events is higher than from data.

\label{sec:results}
\begin{figure}
\includegraphics[width=0.45\textwidth]{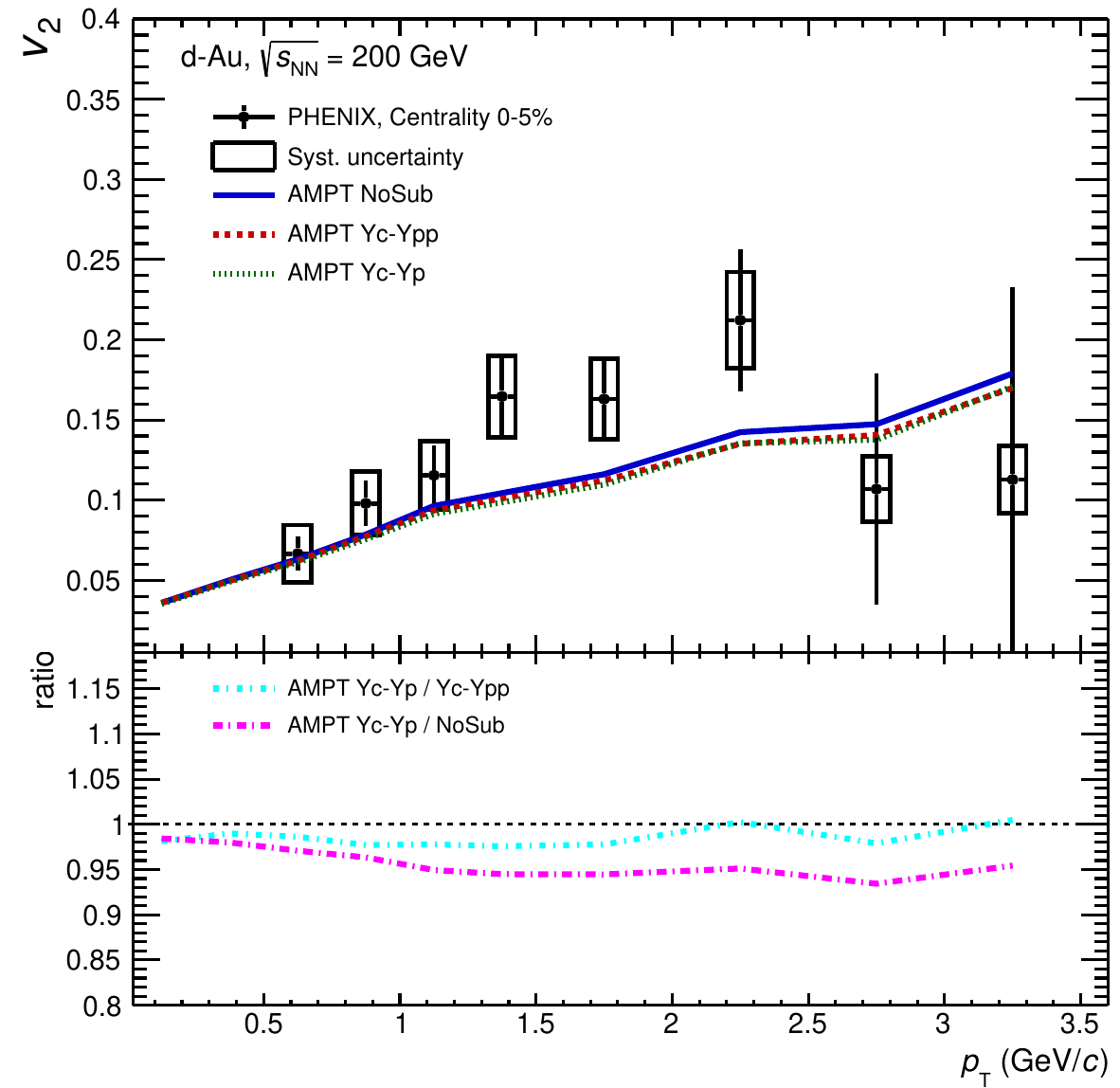}
\caption{In upper panel, the charged particles elliptic flow $v_{2}$ of the 0--5\% most central d-Au collision excess as a function of associated particle $p_{\rm T}$.
The blue solid line is for $v_{2}$ of the 0--5\% most central d-Au collisions with nonflow effects.
The nonflow effects for 0--5\% most central d-Au collisions
reduced by subtracting the per-trigger yield distribution
in peripheral d-Au collisions (green dotted line) or pp collisions (red dashed line).
The $v_{2}$ of the central d-Au collision in the data is
also shown (black circles) from Ref.~\cite{Adare:2013piz}, which obtained under the assumption of factorization:
$c_{2}\left( p^{t}_{\rm T}, p^{a}_{\rm T} \right)
= v_2\left(p^{t}_{\rm T}\right) v_2\left(p^{a}_{\rm T}\right)$.
In bottom panel, the ratio between charged particles $v_{2}$ of most central collisions
reduced by subtracting the per-trigger yield distribution
in peripheral d-Au collisions and pp collisions is
shown by cyan shortdash-dotted line, the ratio between $v_{2}$ of most central d-Au collisions
reduced by subtracting the per-trigger yield distribution
in peripheral d-Au collisions and without nonflow subtracting is shown by pink longdash-dotted line.
}
\label{fig:Azimuthal_conditional_yields_4}
\end{figure}

\label{sec:results}
\begin{figure}
\includegraphics[width=0.49\textwidth]{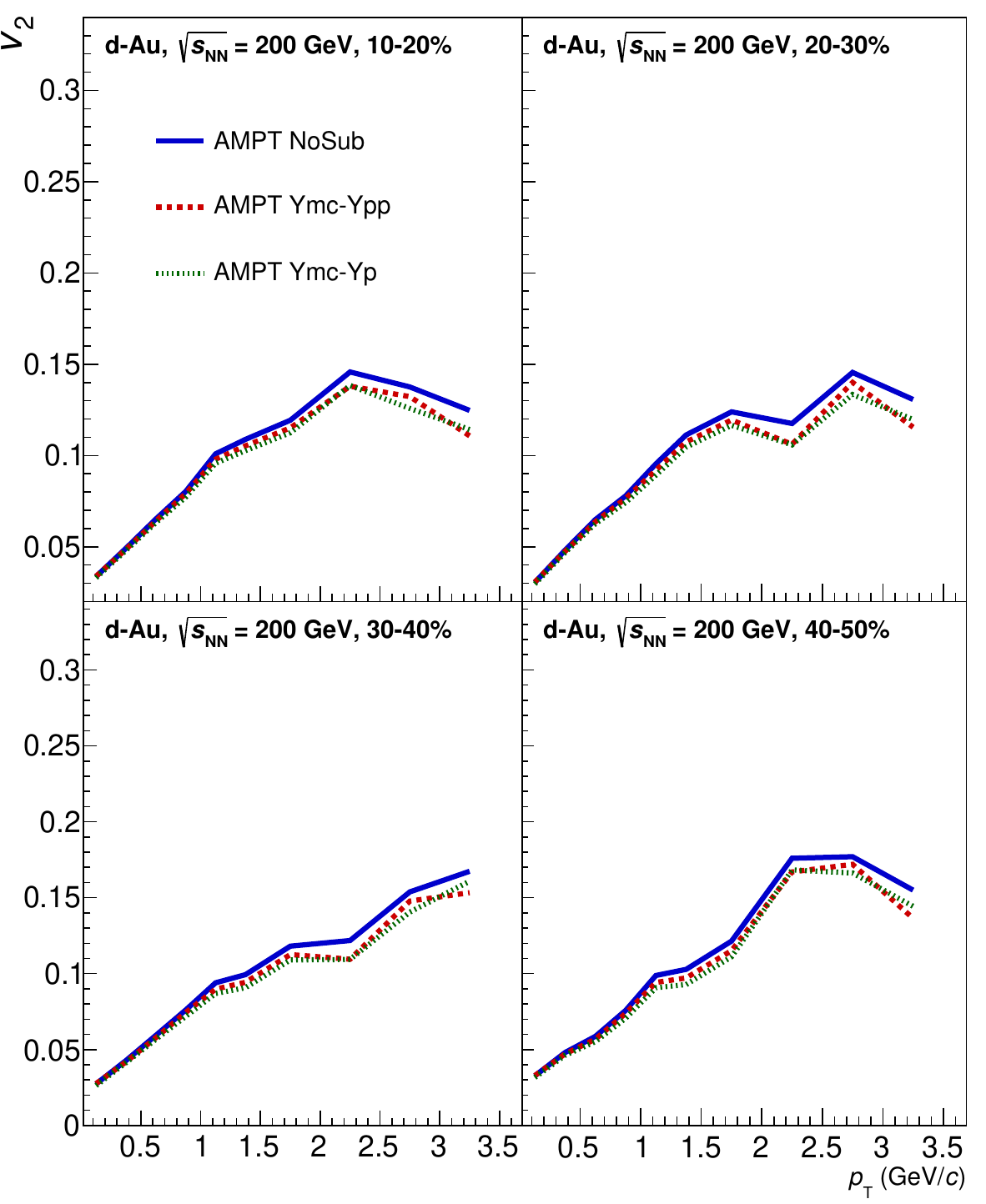}
\caption{The charged particles elliptic flow $v_{2}$ of the 10--20\%,  20--30\%, 30--40\% and 40--50\% collision excess as a function of associated particle $p_{\rm T}$.
The blue solid line is for charged particles $v_{2}$ of the different centrality collisions with nonflow effects.
The nonflow effects
reduced by subtracting the per-trigger yield distribution
in peripheral d-Au collisions (green dotted line) or pp collisions (red dashed line).
}
\label{fig:Azimuthal_conditional_yields_5}
\end{figure}

  From Eq.~(\ref{eq:v2}),charged particles $v_2$ is shown as a function of associated $p_{\rm T}$ in Fig.~\ref{fig:Azimuthal_conditional_yields_4} for central (0--5\%) d-Au collisions.
  We can observe the charged particles $v_2$ value increase with $p_{\rm T}$ distribution,
  this is because at higher $p_{\rm T}$, the particles are more affected by reflecting the initial spatial anisotropy of the collision geometry.
  With the nonflow effect, the blue solid line is higher than red dashed line or green dotted line (without the nonflow effect). The nonflow effects for 0--5\% most central d-Au collisions
  reduced by subtracting the per-trigger yield distribution
  in peripheral d-Au collisions (green dotted line) or pp collisions (red dashed line) are very similar. This could indicate that the nonflow effects (such as jet correlations and resonance decays) in the 0-5\% most central d-Au collisions are not strongly dependent on the subtracting the per-trigger yield distribution in peripheral d-Au collisions
  or pp collision system.
  The charged particles $v_2$ from AMPT events are in agreement with data from 0.5 to 1.2~GeV/$c$, then $v_2$ from AMPT events is lower than data from 1.2 to 2.5~GeV/$c$.

  In bottom panel of Fig.~\ref{fig:Azimuthal_conditional_yields_4} , we can observe the ratio between charged particles $v_{2}$ of most central d-Au collisions
  reduced by subtracting the per-trigger yield distribution
  in peripheral d-Au collisions and pp collisions is
  shown by cyan shortdash-dotted line, the maximum deviation is less than 2\%.
  The ratio between $v_{2}$ of most central d-Au collisions
  reduced by subtracting the per-trigger yield distribution
  in peripheral d-Au collisions and without nonflow subtracting is shown by pink longdash-dotted line, the ratio value at higher $p_{\rm T}$ is larger than at lower $p_{\rm T}$ and
  the maximum deviation is less than 6\%.
  It is worth noting that the nonflow contribution can vary with $p_{\rm T}$ due to different physical mechanisms that contribute to particle production in different $p_{\rm T}$ ranges. For example, at low $p_{\rm T}$, the nonflow contribution may be dominated by resonance decays, while at high $p_{\rm T}$, the nonflow contribution may be dominated by jet-like correlations. Therefore, studying the nonflow contribution ratio as a function of $p_{\rm T}$ can provide important insights into the underlying physics of particle production in heavy-ion collisions.

  The charged particles $v_2$ is shown as a function of associated $p_{\rm T}$ in Fig.~\ref{fig:Azimuthal_conditional_yields_5} at 10--20\%,  20--30\%, 30--40\% and 40--50\%  d-Au collision centrality intervals. We can find the nonflow effects in the mid-central d-Au
  collisions are not strongly dependent on the subtracting the per-trigger yield distribution in peripheral d-Au collisions
  or pp collision system.
  On other hand, we observe the nonflow effects at different collision centrality intervals
  reduced by subtracting the per-trigger yield distribution in peripheral d-Au collisions (green dotted line) or pp collisions (red dashed line) are very similar.

\section{Summary}

  In this work, a multiphase transport model(AMPT) has been used to comprehensively study the behavior of elliptic flow($v_{\rm 2}$) for d-Au collisions at $\sqrt{s_{\rm NN}} = 200$~GeV.
  Two-particle angular correlations of charged particles have been calculated in d-Au collisions at $\sqrt{s_\mathrm{NN}} = 200$~GeV and expressed as associated yields per trigger particle. The Fourier coefficient $v_2$ was extracted from these correlations and studied as a function of $p_{\rm T}$.
  The nonflow effects for central/mid-central d-Au
  collisions at $\sqrt{s_{\rm NN}} = 200$~GeV reduced by subtracting the per-trigger yield distribution in peripheral d-Au collisions at $\sqrt{s_{\rm NN}} = 200$~GeV or pp collisions at $\sqrt{s}=200$~GeV.
  Both techniques give compatible results.
  Discussions about comparisons with measurements from d-Au collisions at $\sqrt{s_{\rm NN}} = 200$~GeV are included.
  We can find the nonflow effects (such as jet correlations and resonance decays) in the central and mid-central d-Au
  collisions are not strongly dependent on the subtracting the per-trigger yield distribution in peripheral d-Au collisions
  or pp collision system.
  By subtracting the per-trigger yield distribution in peripheral d-Au collisions and without nonflow subtraction, the ratio of $v_{2}$ for most central d-Au collisions shows a larger value at higher $p_{\rm T}$ than at lower $p_{\rm T}$. The maximum deviation is less than 6\%.
  By analyzing the nonflow contribution ratio as a function of $p_{\rm T}$, we can gain valuable insights into the underlying physics of particle production in heavy-ion collisions. At low $p_{\rm T}$, resonance decays may dominate the nonflow contribution, whereas at high $p_{\rm T}$, jet-like correlations may be the main contributor.
  On the other hand, we have observed nonflow effects in different collision centrality intervals. These effects have been reduced by subtracting the per-trigger yield distribution in peripheral d-Au collisions or pp collisions, which are remarkably similar.
  Therefore, our work provides a reference for researchers looking into nonflow contribution subtraction in small collision system experiments.

  The authors appreciate the referee for his/her careful reading of the paper and valuable comments. This work is supported in part by the Natural Science Foundation of Henan Province (No.212300410386), Key Research Projects of Henan Higher Education Institutions (No.20A140024), and
  NSFC Key Grant 12061141008, the Scientific Research Foundation of Hubei University of Education
  for Talent Introduction (No. ESRC20220028 and No. ESRC20230002).


\end{document}